# Ultraviolet-ozone treatment: an effective method for fine-tuning optical and electrical properties of suspended and substrate-supported MoS$_2$

Fahrettin Sarcan[1, 2,] (✉), Alex J. Armstrong[1], Yusuf K. Bostan[2,3], Esra Kus[2], Keith McKenna[1], Ayse Erol[2], Yue Wang[1](✉)

[1] *School of Physics, Engineering and Technology, University of York, York, YO10 5DD, United Kingdom*
[2] *Department of Physics, Faculty of Science, Istanbul University, Vezneciler, Istanbul, 34134, Turkey*
[3] *Institut d'Electronique, Microélectronique & Nanotechnologie IEMN CNRS UMR 8520, Université Polytechnique Hauts de France, Valenciennes 59313, France*

Corresponding Authors: Fahrettin Sarcan, fahrettin.sarcan@istanbul.edu.tr ; Yue Wang, yue.wang@york.ac.uk

## ABSTRACT

Ultraviolet-ozone (UV-O$_3$) treatment is a simple but effective technique for surface cleaning, surface sterilization, doping and oxidation, and is applicable to a wide range of materials. In this study, we investigated how UV-O$_3$ treatment affects the optical and electrical properties of molybdenum disulfide (MoS$_2$), with and without the presence of a dielectric substrate. We performed detailed photoluminescence (PL) measurements on 1-7 layers of MoS$_2$ with up to 8 minutes of UV-O$_3$ exposure. Density functional theory (DFT) calculations were carried out to provide insight into oxygen-MoS$_2$ interaction mechanisms. Our results showed that the influence of UV-O$_3$ treatment on PL depends on whether the substrate is present, as well as the number of layers. The PL intensity of the substrate-supported MoS$_2$ decreased dramatically with the increase of UV-O$_3$ treatment time and was fully quenched after 8 mins. However, the PL intensity of the suspended flakes was less affected. 4 minutes of UV-O$_3$ exposure was found to be optimal to produce p-type MoS$_2$, while maintaining above 80% of the PL intensity and the emission wavelength, compared to pristine flakes (intrinsically n-type). Our electrical measurements showed that UV-O$_3$ treatment for more than 6 minutes not only caused a reduction in the electron density but also deteriorated the hole-dominated transport. It is revealed that the substrate plays a critical role in the manipulation of the electrical and optical properties of MoS$_2$, which should be considered in future device fabrication and applications.

## KEYWORDS

MoS$_2$, ultraviolet-ozone (UV-O$_3$), doping, surface treatment, transition metal dichalcogenide, field effect transistor, photoluminescence, density functional theory (DFT)

## 1 Introduction

In recent years, semiconducting two-dimensional transition metal dichalcogenides (2D-TMDs) have become a group of highly demanded materials for next-generation optoelectronic devices such as photodetectors and light emitters [1–4], because of their unique optical, electronic, and structural properties. Despite many advantages of these semiconducting materials, there are some drawbacks such as low carrier concentration and mobility, which result in low electrical conductivity [5,6] compared to the materials already widely used in electronic/optoelectronic technologies, such as Si and GaAs. To realize high-performance optoelectronic devices, both n-type and p-type semiconducting 2D-TMDs are required [7,8]. Conventional doping techniques used for semiconductors are not suitable for 2D materials because they modify their crystal structures and result in a significant deterioration of their optoelectrical properties [9]. On the other hand, thanks to the atomic thickness of 2D-TMDs, their optical and structural properties, as well as carrier dynamics, can be efficiently engineered by different post-growth methods [10]. Surface charge transfer-based doping [11], substitutional doping [12], interstitial doping [13], and vacancy-based doping [14] are the main post-growth doping mechanisms for 2D-TMDs [15]. Based on these mechanisms, there are several reported doping techniques such as chemical treatment [13], ion implantation [16], plasma doping [17], thermal annealing [18], electron beam irradiation [19] and ultraviolet-ozone (UV-O$_3$) treatment [20], which aim to achieve n- and/or p-type doping. Due to the high sensitivity of 2D materials' properties, the main challenge in post-growth doping is to maintain their superb optoelectrical properties for device applications, while controlling the doping concentration with consistency. In this study, we focus on the effect of UV-O$_3$ treatment on the optical and electrical properties of MoS$_2$, which leads to p-type doping of the TMD material.

UV-O$_3$ treatment is a useful process for a wide range of materials for surface modification [21]. It has been employed as an effective tool for defect engineering and doping in graphene and 2D TMDs. There are a few studies in the literature on UV-O$_3$-induced p-type doping on graphene [22–24]. Liang *et al.* presented controllable p-type doping in a range of semiconducting TMDs (MoTe$_2$, WSe$_2$, MoSe$_2$ and PtSe$_2$) and proposed three mechanisms for UV-O$_3$-induced hole doping: (1) charge transfer due to the interaction with oxygen molecules, (2) isoelectronic substitution of chalcogen atoms with oxygen atoms, and (3) charge transport over the oxide surface due to the transition metal oxides formation (MoO$_3$, WO$_3$ etc.) [20]. Zheng *et al.* showed that UV-O$_3$ treatment is an effective method for p-type doping of MoTe$_2$ field-effect transistors and it enhances its electrical performance enormously. The hole concentration and mobility are enhanced by nearly two orders of magnitude, and the conductivity by 5 orders of magnitude [25]. These promising studies are focused on the effect of UV-O$_3$ treatment on the electrical properties of TMDs only. In 2D-TMD optoelectronic devices, the effect of UV-O$_3$ treatment on their optical properties is also critical. There are only a few studies about the effect of UV-O$_3$ treatment on the optical properties of the semiconducting TMDs. Yang *et al.* reported that PL intensity of pristine exfoliated MoS$_2$ decreased and was eventually quenched as UV-O$_3$ exposure time increased from 0 to 10 minutes, which was attributed to the structural degradation [26]. The quenching of PL was also observed on chemical vapour deposited and exfoliated monolayer MoS$_2$ [27] and on exfoliated monolayer WS$_2$ and WSe$_2$ as a result of 6 minutes of UV-O$_3$ treatment due to oxidation [28]. On the other hand, Zheng et al. showed a 37-fold increment in the PL intensity, by converting trilayer MoSe$_2$ to monolayer, with 7 minutes of UV-O$_3$ treatment [29]. While the effect of UV-O$_3$ on the electrical properties is consistent from one study to another, there is a contradiction about the effect of UV-O$_3$ treatment on the optical properties of 2D-TMDs. More importantly, there is no study on the effect of UV-O$_3$ treatment on suspended TMDs. In this paper, we reveal that the effect of UV-O$_3$ treatment does not only depend on the type of materials but also the number of layers and the substrate (the environment). To investigate the layer and substrate dependency of the UV-O$_3$ treatment on the optical properties of semiconducting TMDs, we fabricated suspended and substrate-supported MoS$_2$ samples with different number of layers from monolayer to 7 layers. PL spectroscopy was carried out to investigate the effect on their optical properties with a range of UV-O$_3$ exposure time. Density functional theory (DFT) calculations were performed to understand which oxygen-related mechanism during the UV-O$_3$ exposure could cause p-type doping *via* charge trapping, as well as the role of intrinsic sulphur defects. Further electrical characterisation was performed on a 4L-MoS$_2$ transistor to investigate how the carrier dynamics can be modified by the UV-O$_3$ treatment.

## 2 Experimental

E-beam lithography was used to pattern the squares (4 μm × 4 μm) on Polymethyl methacrylate (PMMA) coated Si$_3$N$_4$ (150 nm)-on-SiO$_2$ substrates. Patterned squares were etched 300 nm in depth with a Reactive Ion Etcher, using a mixture gas of CHF$_3$ and O$_2$. Resist was removed with a resist remover (1165) and acetone, and the substrate was rinsed in isopropyl alcohol (IPA) before drying. Bulk single-crystal MoS$_2$ was purchased from HQ Graphene, which was intrinsically n-doped. Scotch tape and polydimethylsiloxane (PDMS) assisted mechanical exfoliation method was used to obtain MoS$_2$ flakes with different numbers of layers. The flakes larger than 10 μm × 10 μm were transferred either on the etched squares to be suspended or the flat area of the Si$_3$N$_4$-on-SiO$_2$ substrates. The PL spectra of the flakes were measured using a microPL setup equipped with an Andor iDus detector and a 532 nm excitation laser, The laser beam was focused to be a spot of ~1.5 μm in diameter, fitting inside the suspended areas.

DFT calculations were undertaken on 4-layer MoS$_2$, using a plane wave basis set, as implemented in the VASP package [30–32]. A gamma-centered Monkhorst-Pack grid of 2x2x1 *k* points was used to sample the Brillouin zone for all calculations, with geometry optimizations performed to a force tolerance of 0.01 eV/Å. Ideal values of D3 parameters for HSE06 are still an open area of research and could therefore not be sourced from literature, so in this study, we used parameters quoted for the related hybrid functional PBE0 for calculations. Version 5.2 PBE plane wave potentials (PAW) were used for all calculations, with a plane wave cut-off value of 520 eV.

Preliminary geometry optimizations were undertaken with the Perdew-Burke-Ernzerhof (PBE) functional [33], including Grimme's D3 van der Waals corrections [34] to determine the lowest energy adsorption site and adsorbate bond orientations. Further calculations of the band structures, including substitutional defects, were undertaken using the HSE06 functional with D3 corrections, ensuring more accurate energy levels to fully capture any charge-trapping levels [35]. These calculations were initially optimized with one extra electron added to capture the geometry of the charged structure, then reoptimized for the neutral case. This enabled Bader charge analysis to be conducted on both neutral and charged cases to identify if the charged case differed greatly at the defect site, which would be indicative of charge trapping. To reduce the defect density and prevent bands forming from the defect levels, the unit cell used was double the size of the primitive cell in the *a* and *b* directions. A vacuum gap of 20 Å was employed to minimize interaction between periodically repeated slabs.

The band structure was also calculated for a pristine 4-layer MoS$_2$ system at the HSE06 level, with the parameters described above and a high symmetry *k*-point path generated by the sumo package [36]. This was necessary to determine where the defect levels were relative to these bands.

Defect formation energies were calculated for each defect as a function of the chemical potential of sulphur to determine which defects were most stable under different conditions - from sulphur-poor to sulphur-rich. These energies were calculated using Equation (1), where $\mu_i$ are the chemical potentials of species removed or added to the system and $n_i$ is the change in the number of these atoms.

$$E_f^{def} = E^{def} - E^{bulk} - \sum n_i \mu_i \quad (1)$$

Both sulphur-rich and sulphur-poor limits were considered at a constant oxygen chemical potential – with only 2 points necessary due to the clear linear nature of these energies with respect to one chemical potential. These limits were defined as the sulphur chemical potential when $\mu_s = 1/2 E_{S2}$ and $\mu_{Mo} = E_{Mo}$ for the sulphur-rich and sulphur-poor cases respectively, as either of these conditions uniquely defines both $\mu_s$ and $\mu_{Mo}$ according to Equation (2).

$$E_{MoS2} = \mu_{Mo} + 2\mu_S \quad (2)$$

All energies used were calculated at PBE+D3 level of theory and the formation energies for O$_2$ defects were halved to calculate per oxygen atom and ensure these energies are comparable between the different systems.

Electronic states with enhanced localization on oxygen-related defects were identified by examining coefficients in the projection of one-electron wavefunctions onto atoms. To ensure the corresponding band energies were consistent between different supercells the lowest energy molybdenum core state in the pristine surface was a reference for aligning electronic structures. The densities of states were calculated on a grid of 2000 points and plotted using the sumo package [36].

In order to monitor the doping effect with different UV-O$_3$ exposure times, a 4L-MoS$_2$ field effect transistor was fabricated. A 4L MoS$_2$ flake was transferred on SiO$_2$/Si wafer and e-beam lithography was used to pattern a Transfer Length Method (TLM) structure on the flake. Cr/Au (10 nm/60 nm) were deposited in a thermal evaporator, followed by a lift-off process. Electrical characterisation was carried out using an Agilent B2902A double-channel source-meter unit.

## 3 Results and Discussion

It is well known that the electronic band structure and energy bandgap of TMDs depend on the number of layers. The number of layers in MoS$_2$ flakes can be precisely determined by the PL peak wavelength/energy. Flakes with different layer numbers were selectively transferred to the pre-patterned Si$_3$N$_4$ substrates. Figure 1a and 1b show a microscope image of the 4L MoS$_2$ flake on Si$_3$N$_4$ substrate, and a schematic drawing of the sample structure, respectively.

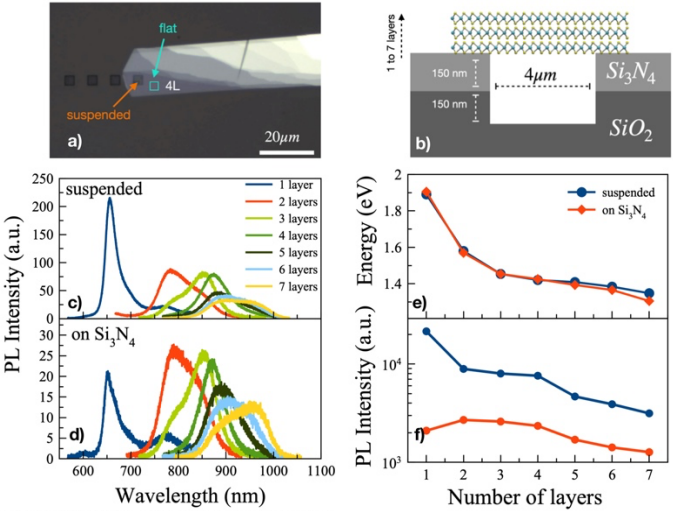

**Figure 1** a) Microscope image of the 4L MoS$_2$ on Si$_3$N$_4$, b) the schematic of the sample, PL spectra of c) suspended and d) supported MoS$_2$ on Si$_3$N$_4$ and e) PL peak energy and f) intensity as a function of layer numbers.

Optical properties of suspended and supported MoS$_2$ flakes on Si$_3$N$_4$ were investigated using micro-PL spectroscopy. As a function of the number of layers (from a single layer to 7 layers) the bandgap of the suspended MoS$_2$ varies from ~1.90 eV to ~1.35 eV (Figure 1e). Up to 45 meV blueshifts of the PL peak energy were observed in the thicker supported flakes, compared to the suspended flakes [37]. There are several studies on the substrate effect on the optical properties of 2D TMDs in the literature [38–40], although most of these studies were focused on the substrate effect on monolayer TMDs. The weaker, broader, and red-shifted PL from the TMDs transferred onto dielectric substrates can be attributed to the amount of moisture at the interface and/or interlayer charge transfer, which results in charge doping [40–42]. In Figure 1, we observe a consistent reduction of PL intensity for all layer numbers, and a small redshift of the peak energy in the thicker supported flakes compared to the suspended flakes. This can be explained by the possibility of more charge transfer in the thicker layers due to their lower energy bandgap [43].

Next, the effect of UV-O$_3$ treatment on suspended and supported MoS$_2$ as a function of number of layers and exposure time has been investigated. Each sample was exposed separately. Figure 2 shows the PL spectra of pristine and UV-O$_3$-treated flakes. As UV-O$_3$ treatment time increases, the PL intensity decreases for all samples. The intensity of the PL emission from the supported MoS$_2$ decreases significantly with UV-O$_3$ treatment time and is eventually fully quenched after 8 mins., while the suspended flakes are less affected by the treatment. Figure 3 shows that there are clear differences in the trend between the suspended and supported flakes. In general, the thinner flakes are more sensitive to UV-O$_3$ treatment for both cases, which is expected. However, for suspended flakes, up to 4 minutes of exposure for all thicknesses does not degrade the PL intensity significantly.

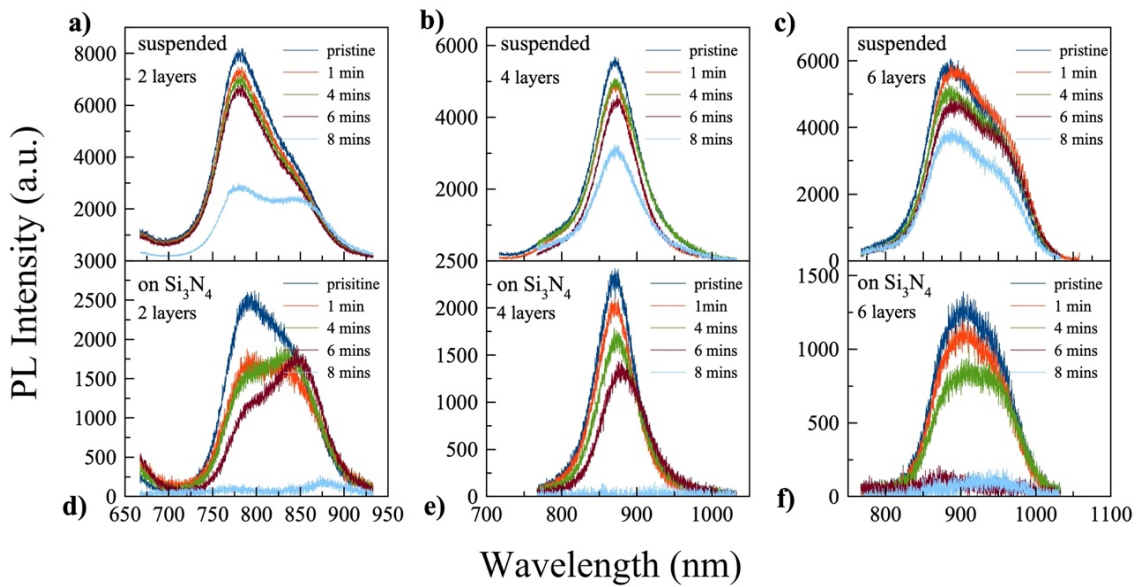

**Figure 2** PL spectra of a) 2 layers, b) 4 layers and c) 6 layers MoS$_2$ flakes with different UV-O$_3$ treatment times: 1, 4, 6, 8 mins.

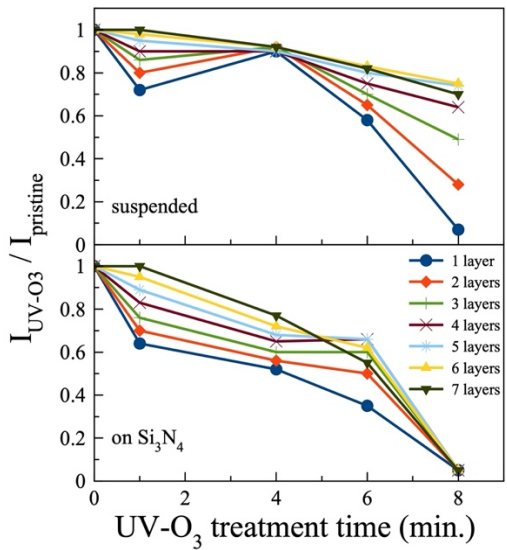

**Figure 3** PL intensity ratio as a function of UV-O$_3$ exposure time.

We use DFT calculations to understand the effect of UV-O$_3$ on suspended TMDs. Conventional UV-O$_3$ cleaners have two dominant UV peaks, at 184 nm and 254 nm. Upon irradiation, molecular oxygen (O$_2$) present in the air is dissociated by radiation at 184 nm. This results in the formation of two radicals of singlet oxygen (O). These radicals continue to react with molecular oxygens forming molecules of ozone (O$_3$). The simulations were carried out for these three species. Sulphur substitutional defects involving these species, as well as surface adsorption onto various non-symmetrically equivalent high symmetry sites, were considered. Both substitution and adsorption of O$_3$ resulted in dissociation into an O defect and a free O$_2$ molecule in the vacuum gap, which is therefore equivalent to the O defects. The formation energies were calculated with a constant oxygen chemical potential of half the energy of an oxygen atom, plotted in Figure 4a, which represents the stability of each type of defect as a function of sulphur chemical potential. In both the sulphur-rich and sulphur-poor extremes, no O$_2$ defects were stable as their formation energies were above zero for all chemical potentials, hence O defects can be concluded to be the most likely defect to form in this system. In the sulphur-poor region, O$_2$/S substitution is the most stable, and for all chemical potentials, O adsorption is the most stable scenario. Figure 4b shows that the unoccupied electronic states introduced by all these defects are higher in energy than the conduction band minimum (CBM) of pristine monolayer MoS$_2$ meaning no PL shift should be induced by the defects. The O$_2$ defects have higher formation energy than the O defects, which indicates that they are likely to be short-lived, but these short-lived O$_2$ defects could be relevant for providing mechanisms to form O defects. In Figure 4c, the excess Bader charge was shown as a function of the vertical position of each atom in the unit cell, in the case of O adsorption and with one added electron. Low localization can be seen at both top and bottom surfaces, with the most localization on the sulphur atom closest to the adsorbed O atom. Even this larger localization is only 0.08e, suggesting this localization is extremely weak and unlikely to trap enough charge to cause p-type doping. One singlet O adsorption between the TMD layers were also investigated at the PBE level (the lower level of theory) for both neutral and +1 electron systems. This gave similar charge densities and defect levels compared to O adsorption on the surface. We also note here that our DFT model reflects the substitution of one sulphur atom in the unit cell (Fig 4c), and the adsorption of one of the three oxygen species (O, O$_2$ and O$_3$) in the suspended 4L MoS$_2$ system; it does not reflect multiple species adsorption or substitution of multiple sulphur atoms, which potentially happens with longer UV-O$_3$ treatment time. The PL results from short time UV-O$_3$ treatment (1 to 4 min) show that no significant reduction in the intensity nor any peak wavelength shift, which is consistent with our DFT results.

Inclusion of the substrate in the DFT calculations is too computationally time-consuming using the hybrid level of theory and requires taking into account extra parameters, such as surface roughness, optical interferences specifically on dielectric substrates, defects, and impurities on the interface etc. which is considered to be future work.

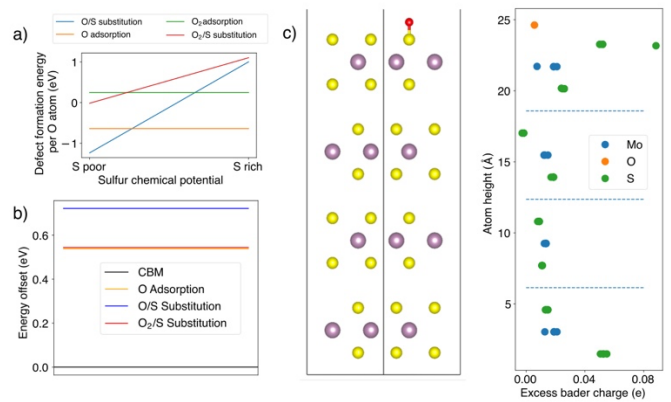

**Figure 4** a) Defect formation energies for O and O$_2$ adsorption and S substitutional defects as a function of sulphur chemical potential; b) Defect energy level offsets for each stable defect mechanism, with respect to the conduction band minimum (CBM) in the pristine 4L-MoS$_2$ system; c) excess Bader charge for each atomic site (right), with sites aligned vertically with unit cell (left), for the O adsorption case with the presence of one added electron.

To understand the effect of UV-O$_3$ on the substrate-supported flakes we analyzed the PL spectra of the 2L and 6L flakes in-depth. The reduction in the PL intensities starts from higher energy on the supported 2L and 6L flakes with up to 6 minutes of UV-O$_3$ treatment (see Figure 2d,f). This can be explained by carrier dynamics involving the presence of the Si$_3$N$_4$ substrate. The bandgap and photogenerated charge carriers of the substrate under UV irritation play a key role in this process. The bandgap of the Si$_3$N$_4$ is ~5 eV. During the UV-O$_3$ exposure, the 184 nm characteristic radiation of the UV source can generate free electrons and holes in the substrate. The generated free electrons diffuse to MoS$_2$, which has a lower energy level than that of the conduction band of Si$_3$N$_4$ in the heterostructure (illustrated in Figure 5a). Such diffusion compensates for the decreased electron density caused by the oxygen adsorption and allows more oxygen defect formation on the MoS$_2$ surface. The transferred electrons also recombine with free holes in MoS$_2$, resulting in non-radiative recombination at the interfaces of the flake and the substrate, hence further reduction of the PL intensity of the supported MoS$_2$. Moreover, we observed an 11 meV redshift of the PL peak energy of the 4L supported flakes, compared to the suspended 4L flakes, which supports the substrate-induced charge carrier dynamics interpretation. Such substrate-dependent carrier dynamics have been observed previously in graphene [24].

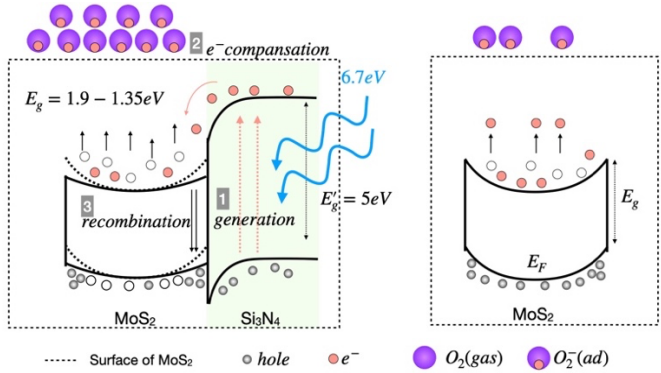

**Figure 5** Illustration of UV-O$_3$ treatment effect on a) supported and b) suspended MoS$_2$.

Finally, we fabricated a substrate-supported field effect transistor with a TLM structure using 4L MoS$_2$ to investigate the doping effect of the UV-O$_3$ treatment. The optical microscope image and illustrations of the fabricated device are presented in Figure 6a. The structure includes 5 transistors with different channel lengths from 2 μm to 4 μm in 0.5 μm step, and the consecutive two electrodes can be used as source and drain. The channel width of the transistors is 12 μm. All transistors (with different channel lengths) exhibit the same trend of the input and output characteristics against UV-O$_3$ treatment. We present the 4 μm length device in detail, in Figure 6 (b-d), as a function of UV-O$_3$ treatment time. Figure 6b shows the input characteristic of the transistor.

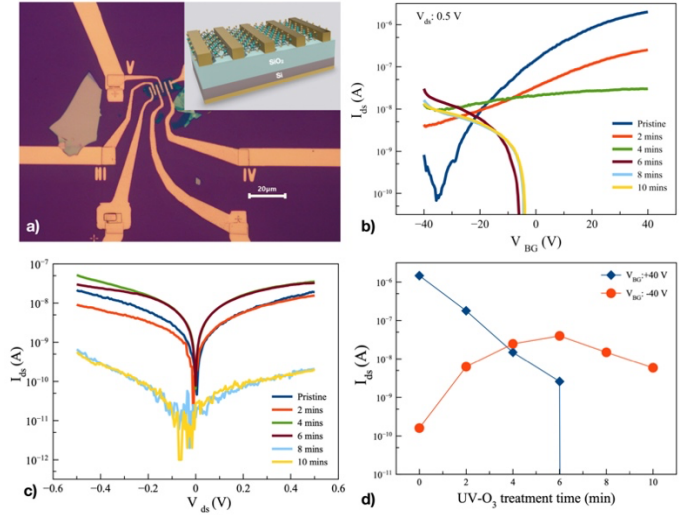

**Figure 6** a) Optical microscope image with 20μm of a scale bar and illustration (inserted) of the MoS$_2$ FET, drain-source current as a function of b) applied gate voltage, c) UV-O$_3$ treatment time and d) applied drain-source voltage.

The pristine MoS$_2$ transistor shows n-type dominant ambipolar characteristics; with increasing the UV-O$_3$ treatment time, the n-type dominancy reduces and eventually, the characteristic changes from electron dominant to hole dominant, as seen in Figure 6b and c. With 6 mins treatment time, the on/off ratio for the n-type dominant reduced from $10^4$ to zero, while that of the p-type dominant one increased from 1 to $10^3$ under 0.5 V bias. On the other hand, above 6 minutes of UV-O$_3$ treatment not only is the electron density reduced but so is the hole density, and the drain-source current is 2 orders of magnitude lower after 8- and 10-minutes of treatment (Figure 6d).

## 4 Conclusion

The effect of UV-O$_3$ treatment on the optical and electrical properties of mono- and multi-layer MoS$_2$ is presented. The PL intensity from substrate-supported MoS$_2$ decreases significantly with UV-O$_3$ treatment time and is fully quenched after 8 minutes of treatment. The PL of suspended flakes is, however, less affected by the UV-O$_3$ treatment. Our DFT results suggest that no significant charge trapping or PL peak energy shift should be expected from the substitution and adsorption of O, O$_2$ and O$_3$ on the surface of a suspended MoS$_2$ flake. We conclude that the substrate plays a critical role in the UV-O$_3$-induced manipulation of the electrical and optical properties of MoS$_2$. The effect of UV-O$_3$ treatment is also layer-thickness dependent; while the thinner flakes could experience a 90% reduction in the PL intensity, the thicker suspended flakes remain above 60% of their original intensity. Our electrical measurements show that above 6 minutes of UV-O$_3$ treatment not only causes a reduction in the electron density, but it also deteriorates the hole transport. Our work suggests that 4 minutes UV-O$_3$ treatment is the optimum time to produce p-type MoS$_2$ and maintain above 80% of PL intensity without any shift in emission wavelength. Longer than 4 minutes leads to deterioration of both optical and electrical properties, which should be taken into consideration in future device fabrication processes.

## Acknowledgements

Y.W. acknowledges a Research Fellowship (TOAST, RF\201718\17131) awarded by the Royal Academy of Engineering. F.S. and AE acknowledge the support from the Scientific Research Projects Coordination Unit of Istanbul University (FBA-2023-39412, FBG-2022-38573, FBG-2021-37896) and The Scientific and Technological Research Council of Turkey (TUBITAK) project (121F169). A.J.A. and K.M. acknowledge the grant of computing time on ARCHER2 via our membership of the UK's HEC Materials Chemistry Consortium, which is funded by EPSRC (EP/R029431 and EP/X035859).

## Author Contributions

F.S. and Y.W. fabricated the samples and performed the photoluminescence measurements. A.J.A. performed the DFT calculations supervised by K.M. F.S., Y.K.B. and E.K. fabricated the sample for the electrical characterisations and performed the measurements. F.S., A.J.A., Y.W., K.M. and A.E. analyzed the results. F.S., A.E, K.M. and Y.W. managed various aspects and funded the project. F.S., A.J.A. and Y.W. wrote the manuscript with contributions from all co-authors. Y.W. oversaw the entire project.

## Corresponding Author

* Fahrettin Sarcan: Department of Physics, Faculty of Science, Istanbul University, Vezneciler, 34134, Istanbul, Turkey, orcid.org/0000-0002-8860-4321, email: fahrettin.sarcan@istanbul.edu.tr

* Yue Wang: School of Physics, Engineering and Technology, University of York, Heslington, York, YO10 5DD, United Kingdom, orcid.org/0000-0002-2482-005X, email: yue.wang@york.ac.uk

## Competing Interests
The authors declare no competing interests.

## Data And Code Availability
The authors declare that all the data and code supporting the findings of this study are available within the article, or upon request from the corresponding author.

# References

[1] Fang H, Liu J, Li H, Zhou L, Liu L, Li J, Wang X, Krauss T F and Wang Y 2018 1305 nm Few‐Layer MoTe2‐on‐Silicon Laser‐Like Emission Laser Photonics Rev 12 1800015

[2] Withers F, Pozo-Zamudio O D, Mishchenko A, Rooney A P, Gholinia A, Watanabe K, Taniguchi T, Haigh S J, Geim A K, Tartakovskii A I and Novoselov K S 2015 Light-emitting diodes by band-structure engineering in van der Waals heterostructures Nat Mater 14 301–6

[3] Dong T, Simões J and Yang Z 2020 Flexible Photodetector Based on 2D Materials: Processing, Architectures, and Applications Adv Mater Interfaces 7 1901657

[4] Datta I, Chae S H, Bhatt G R, Tadayon M A, Li B, Yu Y, Park C, Park J, Cao L, Basov D N, Hone J and Lipson M 2020 Low-loss composite photonic platform based on 2D semiconductor monolayers Nat Photonics 14 256–62

[5] Allain A and Kis A 2014 Electron and Hole Mobilities in Single-Layer WSe2 Acs Nano 8 7180–5

[6] Qu D, Liu X, Huang M, Lee C, Ahmed F, Kim H, Ruoff R S, Hone J and Yoo W J 2017 Carrier‐Type Modulation and Mobility Improvement of Thin MoTe2 Adv Mater 29 1606433

[7] Pham V P and Yeom G Y 2016 Recent Advances in Doping of Molybdenum Disulfide: Industrial Applications and Future Prospects Adv Mater 28 9024–59

[8] Ji H G, Solís‐Fernández P, Yoshimura D, Maruyama M, Endo T, Miyata Y, Okada S and Ago H 2019 Chemically Tuned p‐ and n‐Type WSe2 Monolayers with High Carrier Mobility for Advanced Electronics Adv Mater 31 1903613

[9] Allain A, Kang J, Banerjee K and Kis A 2015 Electrical contacts to two-dimensional semiconductors Nat Mater 14 1195–205

[10] Sarcan F, Fairbairn N J, Zotev P, Severs-Millard T, Gillard D J, Wang X, Conran B, Heuken M, Erol A, Tartakovskii A I, Krauss T F, Hedley G J and Wang Y 2023 Understanding the impact of heavy ions and tailoring the optical properties of large-area monolayer WS2 using focused ion beam Npj 2d Mater Appl 7 23

[11] Luo W, Zhu M, Peng G, Zheng X, Miao F, Bai S, Zhang X and Qin S 2018 Carrier Modulation of Ambipolar Few‐Layer MoTe2 Transistors by MgO Surface Charge Transfer Doping Adv Funct Mater 28 1704539

[12] Loh L, Zhang Z, Bosman M and Eda G 2020 Substitutional doping in 2D transition metal dichalcogenides Nano Res 1–14

[13] Iqbal M W, Elahi E, Amin A, Hussain G and Aftab S 2020 Chemical doping of transition metal dichalcogenides (TMDCs) based field effect transistors: A review Superlattice Microst 137 106350

[14] Xu K, Zhao Y, Lin Z, Long Y, Wang Y, Chan M and Chai Y 2017 Doping of two-dimensional MoS2 by high energy ion implantation Semicond Sci Tech 32 124002

[15] Luo P, Zhuge F, Zhang Q, Chen Y, Lv L, Huang Y, Li H and Zhai T 2018 Doping engineering and functionalization of two-dimensional metal chalcogenides Nanoscale Horizons 4 26–51

[16] Chen R, Liu Q, Liu J, Zhao X, Liu J, He L, Wang J, Li W, Xiao X and Jiang C 2019 Design of high performance MoS2-based non-volatile memory via ion beam defect engineering 2d Mater 6 034002

[17] Yang W, Sun Q-Q, Geng Y, Chen L, Zhou P, Ding S-J and Zhang D W 2015 The Integration of Sub-10 nm Gate Oxide on MoS2 with Ultra Low Leakage and Enhanced Mobility Sci Rep-uk 5 11921

[18] Iacovella F, Koroleva A, Rybkin A G, Fouskaki M, Chaniotakis N, Savvidis P and Deligeorgis G 2021 Impact of thermal annealing in forming gas on the optical and electrical properties of MoS2 monolayer J Phys Condens Matter 33 035001

[19] Luo T, Pan B, Zhang K, Dong Y, Zou C, Gu Z and Zhang L 2021 Electron beam lithography induced doping in multilayer MoTe2 Appl Surf Sci 540 148276

[20] Liang Q, Gou J, Arramel, Zhang Q, Zhang W and Wee A T S 2020 Oxygen-induced controllable p-type doping in 2D semiconductor transition metal dichalcogenides Nano Res 13 3439–44

[21] Vig J R 1985 UV/ozone cleaning of surfaces J Vac Sci Technology Vac Surfaces Films 3 1027–34

[22] Li W, Liang Y, Yu D, Peng L, Pernstich K P, Shen T, Walker A R H, Cheng G, Hacker C A, Richter C A, Li Q, Gundlach D J and Liang X 2013 Ultraviolet/ozone treatment to reduce metal-graphene contact resistance Appl Phys Lett 102 183110

[23] Chen C W, Ren F, Chi G-C, Hung S-C, Huang Y P, Kim J, Kravchenko I I and Pearton S J 2012 UV ozone treatment for improving contact resistance on graphene J Vac Sci Technology B Nanotechnol Microelectron Mater Process Meas Phenom 30 060604

[24] Liu L, Cao Z, Wang W, Wang E, Cao Y and Zhan Z 2016 Substrate-dependent resistance decrease of graphene by ultraviolet-ozone charge doping Rsc Adv 6 62091–8

[25] Zheng X, Zhang X, Wei Y, Liu J, Yang H, Zhang X, Wang S, Xie H, Deng C, Gao Y and Huang H 2020 Enormous enhancement in electrical performance of few-layered MoTe2 due to Schottky barrier reduction induced by ultraviolet ozone treatment Nano Res 13 952–8

[26] Yang H I, Park S and Choi W 2018 Modification of the optoelectronic properties of two-dimensional MoS2 crystals by ultraviolet-ozone treatment Appl Surf Sci 443 91–6

[27] Jung C, Yang H I and Choi W 2019 Effect of Ultraviolet-Ozone Treatment on MoS2 Monolayers: Comparison of Chemical-Vapor-Deposited Polycrystalline Thin Films and Mechanically Exfoliated Single Crystal Flakes Nanoscale Res Lett 14 278

[28] Kang M, Yang H I and Choi W 2019 Oxidation of WS2 and WSe2 monolayers by ultraviolet-ozone treatment J Phys D Appl Phys 52 505105

[29] Zheng X, Wei Y, Liu J, Wang S, Shi J, Yang H, Peng G, Deng C, Luo W, Zhao Y, Li Y, Sun K, Wan W, Xie H, Gao Y, Zhang X and Huang H 2019 A homogeneous p–n junction diode by selective doping of few layer MoSe 2 using ultraviolet ozone for high-performance photovoltaic devices Nanoscale 11 13469–76

[30] Kresse G and Hafner J 1993 Ab initio molecular dynamics for liquid metals Phys. Rev. B 47 558–61

[31] Kresse G and Furthmüller J 1996 Efficiency of ab-initio total energy calculations for metals and semiconductors using a plane-wave basis set Comput. Mater. Sci. 6 15–50

[32] Kresse G and Furthmüller J 1996 Efficient iterative schemes for ab initio total-energy calculations using a plane-wave basis set Phys. Rev. B 54 11169–86

[33] Perdew J P and Levy M 1983 Physical Content of the Exact Kohn-Sham Orbital Energies: Band Gaps and Derivative Discontinuities Phys. Rev. Lett. 51 1884–7

[34] Grimme S, Antony J, Ehrlich S and Krieg H 2010 A consistent and accurate ab initio parametrization of density functional dispersion correction (DFT-D) for the 94 elements H-Pu J. Chem. Phys. 132 154104

[35] Peralta J E, Heyd J, Scuseria G E and Martin R L 2006 Spin-orbit splittings and energy band gaps calculated with the Heyd-Scuseria-Ernzerhof screened hybrid functional Phys. Rev. B 74 073101

[36] Ganose A M, Jackson A J and Scanlon D O 2018 sumo: Command-line tools for plotting and analysis of periodic ab initio calculations J. Open Source Softw. 3 717

[37] Mak K F, Lee C, Hone J, Shan J and Heinz T F 2010 Atomically Thin MoS2: A New Direct-Gap Semiconductor Phys Rev Lett 105 136805

[38] Lippert S, Schneider L M, Renaud D, Kang K N, Ajayi O, Kuhnert J, Halbich M-U, Abdulmunem O M, Lin X, Hassoon K, Edalati-Boostan S, Kim Y D, Heimbrodt W, Yang E-H, Hone J C and Rahimi-Iman A 2017 Influence of the substrate material on the optical properties of tungsten diselenide monolayers 2d Mater 4 025045

[39] Buscema M, Steele G A, Zant H S J van der and Castellanos-Gomez A 2014 The effect of the substrate on the Raman and photoluminescence emission of single-layer MoS2 Nano Res 7 561–71

[40] Chae W H, Cain J D, Hanson E D, Murthy A A and Dravid V P 2017 Substrate-induced strain and charge doping in CVD-grown monolayer MoS2 Appl Phys Lett 111 143106

[41] Amani M, Chin M L, Mazzoni A L, Burke R A, Najmaei S, Ajayan P M, Lou J and Dubey M 2014 Growth-substrate induced performance degradation in

chemically synthesized monolayer MoS2 field effect transistors Appl Phys Lett 104 203506

[42] Kung Y, Hosseini N, Dumcenco D, Fantner G E and Kis A 2019 Air and Water‐Stable n‐Type Doping and Encapsulation of Flexible MoS2 Devices with SU8 Adv Electron Mater 5 1800492

[43] Iqbal M W, Shahzad K, Akbar R and Hussain G 2019 A review on Raman finger prints of doping and strain effect in TMDCs Microelectron Eng 219 111152